\documentclass{article}

\PassOptionsToPackage{numbers}{natbib}

 \usepackage[final]{neurips_2025_ml4ps}

\usepackage[utf8]{inputenc} 
\usepackage[T1]{fontenc}    
\usepackage{hyperref}       
\usepackage{url}            
\usepackage{booktabs}       
\usepackage{amsfonts}       
\usepackage{nicefrac}       
\usepackage{microtype}      
\usepackage{xcolor}         
\usepackage{graphicx}
\usepackage{amsmath}
\usepackage{amssymb}
\usepackage{wrapfig}
\usepackage{threeparttable}

\title{Multi-Modal Masked Autoencoders for Learning Image-Spectrum Associations for Galaxy Evolution and Cosmology}

\author{%
  Morgan Himes$^{1}$\\
  \texttt{morganhimes@ucla.edu} \\
  \And
  Samiksha Krishnamurthy$^{2}$\\
  \texttt{samikris24@ucla.edu} \\
  \And
  Andrew Lizarraga$^{3}$\\
  \texttt{andrewlizarraga@ucla.edu} \\
  \And
  Srinath Saikrishnan$^{4}$\\
  \texttt{srinathsai22@ucla.edu} \\
  \And
  Vikram Seenivasan$^{1}$\\
  \texttt{vikrams25@ucla.edu} \\
  \And 
  Jonathan Soriano$^{1}$\\
  \texttt{jonathansoriano@astro.ucla.edu} \\
  \And
  Ying Nian Wu$^{3}$\\
  \texttt{ywu@stat.ucla.edu} \\
  \And 
  Tuan Do$^{1}$\\
  \texttt{tdo@astro.ucla.edu} \\
}

\begin{document}

\maketitle

\begin{center}
$^{1}$ Department of Physics and Astronomy, UCLA, Los Angeles, CA 90095 \\
$^{2}$ Department of Electrical and Computer Engineering, UCLA, Los Angeles, CA 90095 \\
$^{3}$ Department of Statistics and Data Science, UCLA, Los Angeles, CA 90095 \\
$^{4}$ Department of Computer Science, UCLA, Los Angeles, CA 90095 \\
\end{center}

\vspace{1em}

\begin{abstract}
  Upcoming surveys will produce billions of galaxy images but comparatively few spectra, motivating models that learn cross-modal representations. We build a dataset of 134,533 galaxy images (HSC-PDR2) and spectra (DESI-DR1) and adapt a Multi-Modal Masked Autoencoder (MMAE) to embed both images and spectra in a shared representation. The MMAE is a transformer-based architecture, which we train by masking 75\% of the data and reconstructing missing image and spectral tokens. We use this model to test three applications: spectral and image reconstruction from heavily masked data and redshift regression from images alone. It recovers key physical features, such as galaxy shapes, atomic emission line peaks, and broad continuum slopes, though it struggles with fine image details and line strengths. For redshift regression, the MMAE performs comparably or better than prior multi-modal models in terms of prediction scatter even when missing spectra in testing. These results highlight both the potential and limitations of masked autoencoders in astrophysics and motivate extensions to additional modalities, such as text, for foundation models.
\end{abstract}

\section{Introduction}

Next-generation astronomical surveys will image billions of galaxies, producing catalogs thousands of times larger than existing datasets to study how galaxies form and evolve across cosmic time \citep{ivezic_lsst_2019, laureijs2011a}. Galaxy spectra encode physically relevant information, such as redshift ($z$), a measure of how much a galaxy's light has been stretched by the universe's expansion. However, obtaining a spectrum can take roughly 100 times longer than capturing an image. While spectroscopic redshifts measured from a galaxy’s spectrum are the most accurate, astronomers rely on photometric redshifts from images for this reason \citep{newman_photometric_2022}. Since spectroscopy is infeasible at the scale of upcoming surveys, this motivates the development of models that leverage imaging to reconstruct missing spectra.

\paragraph{Related Work}
Machine learning methods have been used in astronomy to address these challenges. Multi-layer perceptrons, convolutional, and Bayesian neural networks (MLP, CNN, BCNN) have been widely applied for photometric redshift estimation \citep[e.g.,][]{lahav2012, carrasco_kind_tpz_2013, bonnett_using_2015, beck_photometric_2016, sadeh_annz2_2016, pasquet_photometric_2019,dey_photometric_2022,schuldt_photometric_2021,soriano2024, jones_redshift_2024}. Additionally, QuasarNET applies CNNs to spectra for redshift estimation while AstroMAE uses masked autoencoders (MAEs) on galaxy images \citep{busca_quasarnet_2018, fathkouhi_astromae_2024}. Diffusion models have been used for generating galaxy images across redshifts \citep{lizarraga2024, fan2025, martinez2024}. Multi-modal efforts include AstroCLIP, which jointly embeds images and spectra for tasks such as classification and redshift estimation \citep{parker_astroclip_2024}. 

\section{Contributions}

We adapt a Multi-Modal Masked Autoencoder (MMAE) \citep{avidan_multimae_2022} for reconstructing spectra and images from incomplete data, an example of multimodal learning with missing modality \citep{wu_deep_2024}. We build a dataset of spectra and images for training and testing with partially or fully masked spectra, reflecting the real-world constraints of imaging surveys. Redshift regression serves as an auxiliary task to demonstrate the utility of the learned representations. We extend redshift regression to $z \sim 4$, substantially exceeding the redshift range ($z \lesssim 0.5$) explored in prior models \citep[e.g.][]{parker_astroclip_2024,fathkouhi_astromae_2024}.

Masked autoencoders have been applied in scientific contexts such as medical imaging \citep{zhou2023}, but their use for multi-modal astronomical data is underexplored. AstroMAE \citep{fathkouhi_astromae_2024} employs the use of an MAE for redshift regression on galaxy images, but our work is, to our knowledge, the first instance of using an MAE for joint multi-modal reconstruction on images and spectra in astronomy. 

\section{Data}
\label{sec:data}

We assembled a multi-modal dataset\footnote{Available at \href{https://zenodo.org/uploads/16989593}{Zenodo (access link)}} (referred to as GalaxiesML-Spectra) of 134,533 galaxies, each with 5-band images, 1D spectra, and spectroscopic redshifts. The maximum redshift is $z=4.119$, providing greater coverage than previous datasets used for similar studies. The images are available in two size options ($64\times64$ or $127\times127$ pixels), but our model was trained using the $64\times64$ version. All galaxy images in the dataset are contained in GalaxiesML, a machine learning image dataset built upon the Hyper-Suprime-Cam (HSC) Survey PDR2, covering g, r, i, z, y bands with a median seeing in the i-band of about 0.6 arcsec \citep{do_galaxiesml_2024, hscpdr2}. GalaxiesML is biased towards lower redshifts and brighter magnitudes, as shown in Table \ref{summary-table}. Spectra and redshifts were acquired from the DESI Data Release 1 (DR1), the largest spectroscopic galaxy redshift survey to date \citep{collaboration_data_2025}. The spectra span a wavelength range of 3600–9824 Å with a spectral resolution of 2000 (at 3600 Å) to 5500 (at 9800 Å). GalaxiesML was cross-matched with DESI DR1. Stars and sources with duplicates, inconsistent redshifts, or >80\% missing spectral data were removed, reducing the cross-matched dataset from 136,939 sources to 134,533. Spectral artifacts were removed and replaced with the median flux. A summary of the datasets is available in Table \ref{summary-table}. Our released dataset contains unaugmented spectra after quality cuts and artifact removal. Data augmentation is performed as described in Section \ref{sec:methods}. 

\begin{table}
  \centering
  \begin{threeparttable}
    \caption{Data summary.}
    \label{summary-table}
    \begin{tabular}{lrrrrl}
      \toprule
      Dataset & Data Type & Source Count & $z$ (90th pct.) & $z$ (max) & i-mag (90th pct.) \\
      \midrule
      DESI DR1 & Spec, $z$ & 20,283,824 & 1.343 & 6.857 & -- \\
      GalaxiesML & Img & 286,401 & 1.155 & 4.000 & 22.171 \\
      GalaxiesML-Spectra & Spec, Img, $z$ & 134,533 & 1.581 & 4.119 & 20.635 \\
      \bottomrule
    \end{tabular}
    \begin{tablenotes}
      \footnotesize
      \item Note: Disagreement between HSC and DESI spectroscopic redshifts causes a discrepancy in $z$ (max).
    \end{tablenotes}
  \end{threeparttable}
\end{table}

\section{Methods}
\label{sec:methods}

Our MMAE model\footnote{Code available with \href{https://github.com/astrodatalab/himes_2025/}{GitHub (access link)}} is a transformer-based architecture that models images and spectra to perform image and spectrum reconstruction and redshift regression. A patch-based tokenization strategy is used for both modalities. We follow the Vision Transformer (ViT) \citep{dosovitskiy2021} formulation to convert images into patch tokens. With a 2D convolution, galaxy images of shape $64\times 64 \times 5$ are divided into $8\times8\times 5$ patches and projected into a 256-dimensional embedding. A 2D learnable positional embedding of dimension 256 is added to preserve spatial ordering. Before passing spectra into the model, we perform data augmentation by normalizing as described in \citep{busca_quasarnet_2018} and downsampling from 7783 to 259 flux pixels. An analogous 1D patch embedding process is applied for spectra, using patches of length 8 and a linear projection. 

A random 75\% subset of image and spectra patch tokens is zeroed out for each galaxy. After masking, each modality is separately encoded using a 1D transformer encoder with depth 4, 8 attention heads, and a dropout rate of 0.1. For modality fusion, we employ four layers of cross-attention blocks in which image features query spectral features and vice versa. This allows spectral structure to guide the interpretation of morphological features, informing the model as it reconstructs missing tokens from unmasked tokens in both modalities. After fusion, the tokens are aggregated via attention pooling to produce global image and spectrum embeddings which are concatenated into a joint representation. 

The joint representation feeds into three task-specific heads: image and spectrum decoders, and a redshift regression head. The decoders are MLPs with GeLU activations \citep{hendrycks2023} and dropout, while the regression head maps the fused embedding to a scalar redshift. Unlike the common practice of performing regression after MAE feature extraction, we integrate redshift prediction directly into masked autoencoding, which is novel in a multi-modal setting.

The dataset is randomly split into 70\% training, 15\% validation, and 15\% testing. In training, 50\% of the spectra are randomly zeroed entirely to avoid overreliance and simulate the real-life missing modality case. We optimize with AdamW \citep{loshchilov2019} (weight decay 0.01, learning rate 0.0001), gradient clipping,  and a loss function that is a weighted sum of three terms: mean-squared error reconstruction losses on masked regions for images and spectra and a redshift loss defined as $\mathcal{L}_z=1-\frac{1}{1+(dz/0.15)^2}$ where $dz=(z_\text{pred}-z_\text{spec})/(1+z_\text{spec})$ and $z_\text{pred}$ and $z_\text{spec}$ are the predicted and spectroscopic redshifts \citep{tanaka2018}. The weights for the image, spectrum, and redshift losses are tunable parameters that have been set at 0.1, 0.01, and 1.0, respectively, to account for the difference in scale of the individual loss curves as identified through preliminary parameter tuning. In further studies, we will explore dynamic weighting with the incorporation of physics-aware terms such as line centers, widths, and ratios, continuum slope metrics, and perceptual or structure losses. The impact of masking probability itself will be explored through a study of model performance as a function of masking ratio, and we will explore physically-motivated masking including bandpass gaps and pixel-level instrument errors.

\section{Results and discussion}

\subsection{Feature reconstruction}
We evaluated the model's ability to reconstruct spectral features and images on a held-out test set of 20,181 galaxies. For image reconstruction, the model reproduces the shape and color of galaxies but struggles with fine morphological details in nearby galaxies ($z<0.1$) and sky background noise. For spectral reconstruction, the model captures the broad spectral continuum shapes, even in cases where the entire spectrum is masked at test time, but fails to reproduce the random noise in the original spectra. 

\begin{figure}
    \centering
    \includegraphics[width=0.9\linewidth]{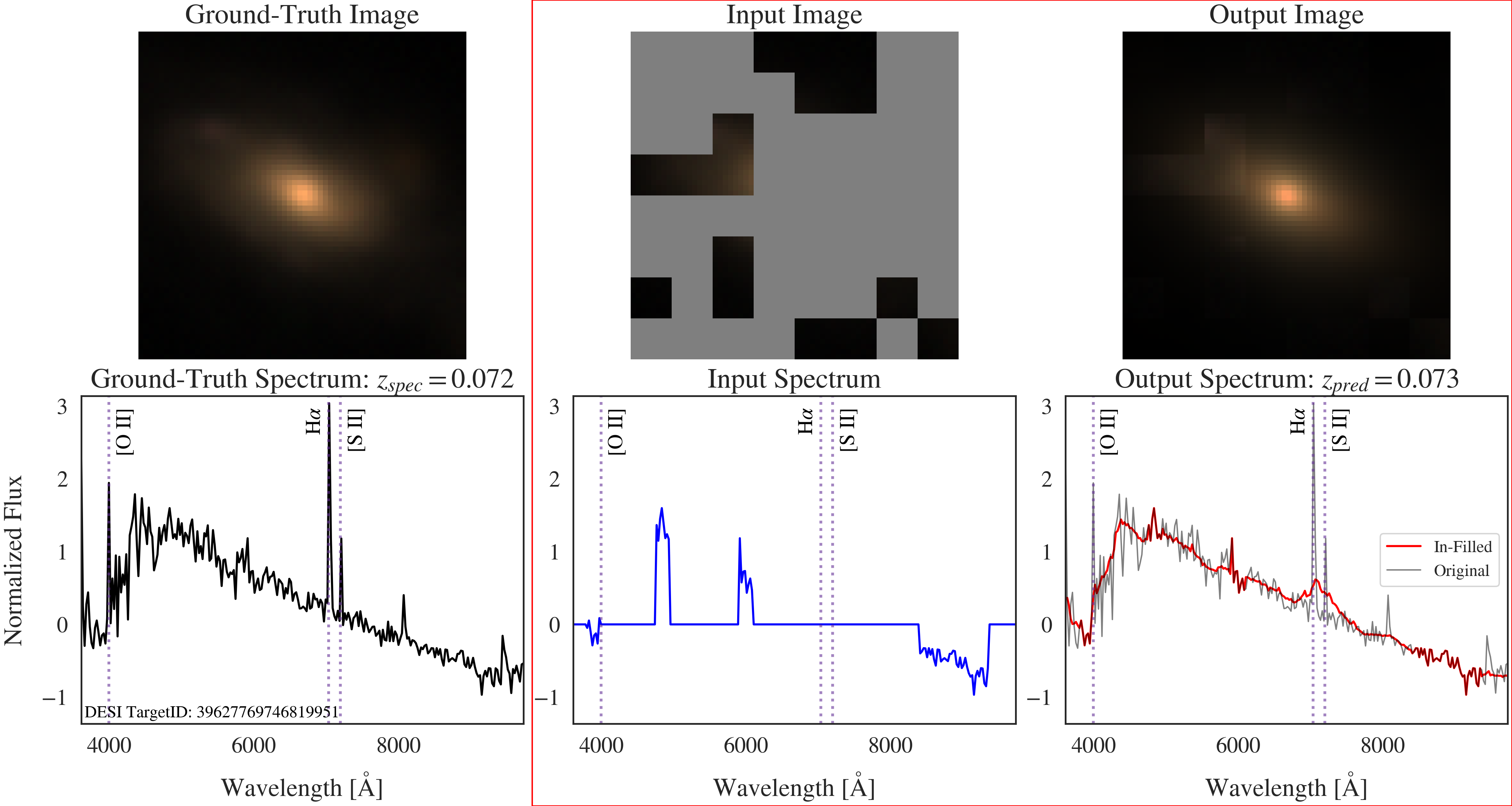}
    \caption{The model's reconstruction process is shown for a low redshift source with 75\% masking of both modalities. We measure the peak location, amplitude, and width of H-$\alpha$ in the augmented and generated spectra. The H-$\alpha$ line has an observed center at 7042.8 Å with a height of 3.04 and a width of 34.5 Å, while the model reconstructed it at 7066.8 Å with a height of 0.62 and a width of 528 Å.}
    \label{fig:lowredshift}
\end{figure}

\begin{figure}[ht]
    \centering
    \includegraphics[width=0.9\linewidth]{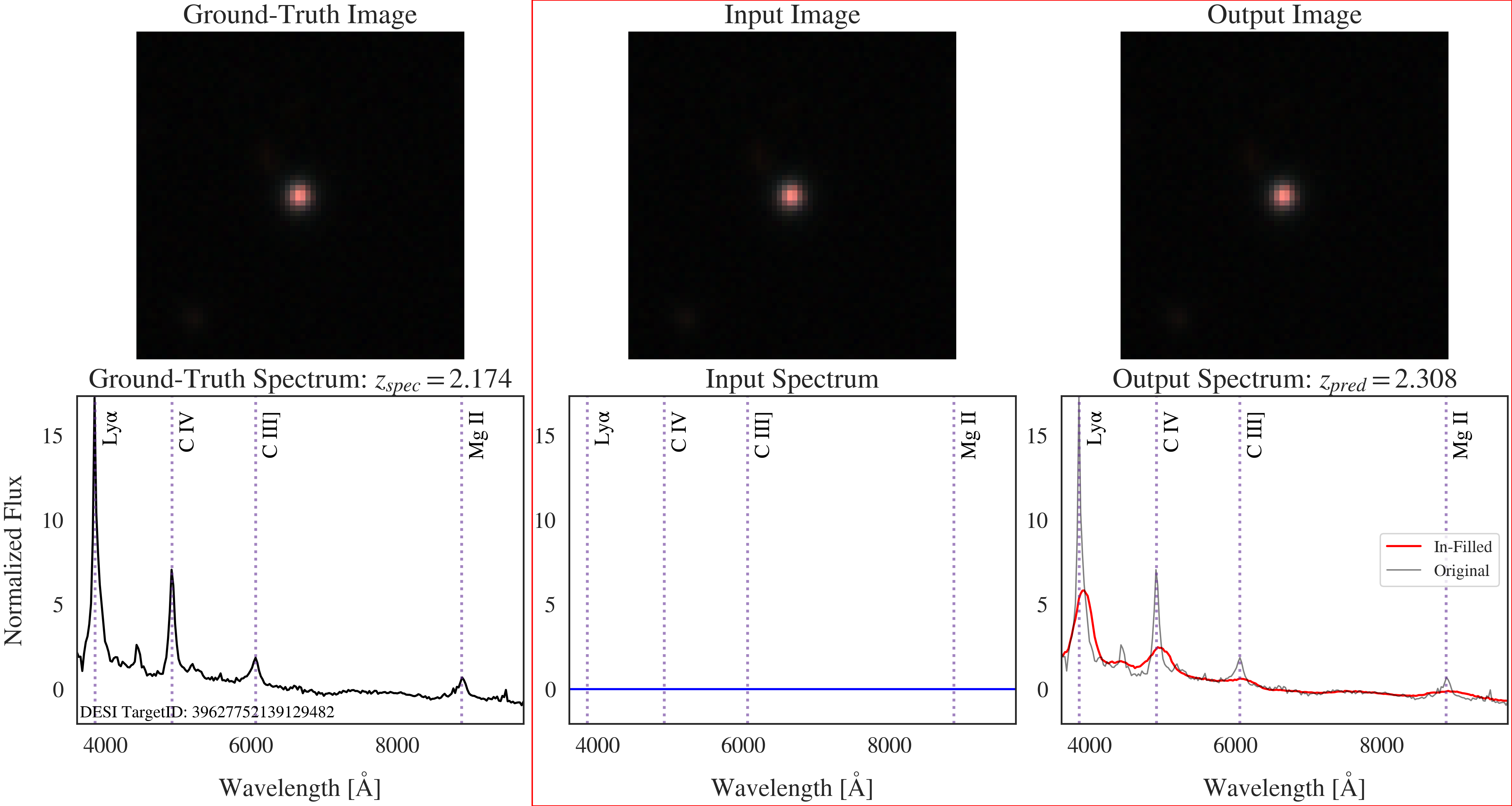}
    \caption{The model's reconstruction process is shown for a high redshift source with a fully masked spectrum and a fully unmasked image. We measure the peak location, amplitude, and width of Lyman-$\alpha$ and C IV in the augmented and generated spectra. The Lyman-$\alpha$ line has an observed center at 3851.6 Å with a height of 17.24 and a width of 48 Å, compared to a reconstructed center at 3923.6 Å, height 5.84, and width 312 Å. Similarly, the C IV line has an observed center at 4907.6 Å, height 7.07, and width 72 Å, while the reconstructed line is at 4931.6 Å, with height 2.48 and width 648 Å.}
    \label{fig:highredshift}
\end{figure}

The model reproduces the locations of some common emission lines. For low-redshift ($z\sim 0.1$) galaxies, it roughly reproduces the position of the H-$\alpha$ (6563 Å rest frame) emission line, as in Figure \ref{fig:lowredshift}. At high redshift ($z\sim 2)$, the model reconstructs the location of Lyman-$\alpha$ (1215.67 Å rest frame) and, in some cases, C IV (1549 Å rest frame); see Figure \ref{fig:highredshift}. In all cases, line widths are systematically overestimated and the heights are underestimated. Reconstruction of other emission lines is poor, and the model occasionally wrongly generates Lyman-$\alpha$ emission for lower redshift, compact galaxies. The model does not capture the ratios of line fluxes which is a diagnostic tool used by astronomers to determine physical properties \citep{bpt1981}. While the MMAE has learned to capture particularly common lines and global structure, spectrum generation remains challenging for weaker and less common emission lines or ambiguous morphology. Further studies will be conducted to assess model's ability to reproduce masked data using quantitative reconstruction metrics.

\subsection{Redshift regression}

\begin{figure}
  \centering
  \includegraphics[width=0.7\linewidth]{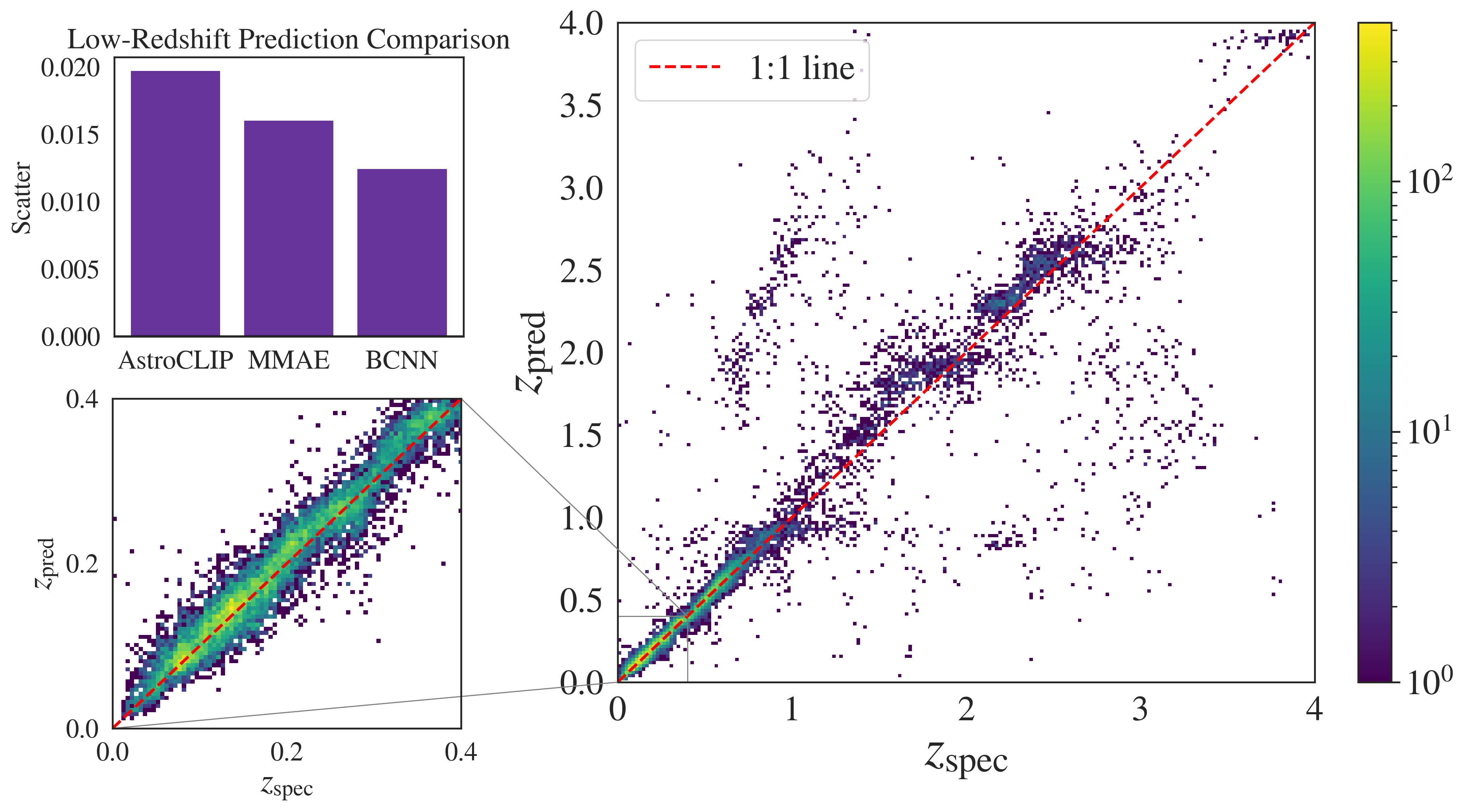}
  \caption{The model's redshift regression results for the entire redshift range are shown (right). The redshift predictions were obtained from test data that had 25\% of the image masked and 100\% of the spectrum masked. The low-redshift regime used for comparison to AstroCLIP \citep{parker_astroclip_2024} is shown in more detail in the bottom left. The top left panel shows the scatter of the MMAE compared to AstroCLIP and a BCNN model \citep{jones_redshift_2024} for this low-redshift regime. Lower scatter corresponds to more precise predictions.}
  \label{fig:redreg}
\end{figure}

Redshift regression was performed on the same test set. The model was provided with a 25\% masked galaxy image and a fully masked spectrum, reflecting the observational constraints posed by upcoming surveys, which will not have spectra for the vast majority of galaxy images. Notably, masking 25\% of the image produces better overall redshift regression results than supplying the model with the entire image. We suspect this is because slight masking serves as a form of regularization, preventing the model from over-fitting to small-scale features. The model achieves higher accuracy at lower redshift ($z\lesssim 1)$ but degrades for higher redshift sources; we attribute this to the limited amount of high redshift data which restricts the model's ability to generalize. We intend to incorporate additional high redshift sources with imaging from the DESI Legacy Imaging Surveys to supplement this gap \citep{DESILS}. The predicted versus true redshift relation (see Figure \ref{fig:redreg}) exhibits distinct step-like structure possibly corresponding to redshift intervals where strong spectral features shift into or out of the spectrograph's wavelength range, such as Lyman-$\alpha$ at $z\sim2$.

We find that our model achieves better or comparable results to other models using images and spectra for redshift regression in terms of scatter, measured as the normalized median absolute deviation. The spectrum was fully masked in testing to resemble real-world survey conditions. Our MMAE model attains $\sigma_{\mathrm{NMAD}} = 0.016$ when tested with 25\% image masking, which is slightly better than AstroCLIP \citep{parker_astroclip_2024}  ($\sigma_{\mathrm{NMAD}} = 0.020$) in the low-redshift regime ($z \lesssim 0.4$, the limit of the AstroCLIP results available to the authors). This comparison is not entirely equivalent, as AstroCLIP was trained to a contrastive objective of aligning images and spectra and did not optimize for redshift prediction. However, we include this comparison as AstroCLIP is the strongest publicly available multi-modal model utilizing the same two modalities. Furthermore, evaluating on the shared task of image-only redshift prediction demonstrates how incorporating masked reconstruction influences downstream inference even with differing training objectives. Masking 25\% of each image produces better redshift regression results than supplying the model with the entire image, in which case $\sigma_{\mathrm{NMAD}} = 0.026$, no longer out-performing other methods. Additionally, a fine-tuned BCNN \citep{jones_redshift_2024}  achieves lower scatter overall ($\sigma_{\mathrm{NMAD}} = 0.012$), indicating that our transformer-based approach remains less robust. This outcome is consistent with prior work suggesting that transformer-based architectures underperform relative to inception-style convolutional models for redshift prediction \citep{fathkouhi_astromae_2024}.


\paragraph{Conclusion} This work demonstrates the potential of using a masked autoencoder on images and spectra for applications in galaxy evolution and cosmology. However, the current model has key limitations that limit its generalizability, as it struggles to reproduce fine details of images and spectra and shows degraded redshift regression accuracy at higher redshift. These limitations highlight the need for more physically motivated training objectives and robust architectures capable of capturing subtle but astrophysically meaningful features. Future work will incorporate physics-aware loss terms and observationally realistic masking strategies. In addition, this work can be easily extended to additional modalities such as  such as textual metadata and natural language descriptions for foundation models for astronomy. Such models will be essential for scaling to the unprecedented data volumes and observational constraints of upcoming surveys.

\clearpage{}

\begin{ack}

Partial support for this work was provided by the Alfred P. Sloan Foundation and the NSF DGE-2034835.

This research used data obtained with the Dark Energy Spectroscopic Instrument (DESI). DESI construction and operations is managed by the Lawrence Berkeley National Laboratory. This material is based upon work supported by the U.S. Department of Energy, Office of Science, Office of High-Energy Physics, under Contract No. DE–AC02–05CH11231, and by the National Energy Research Scientific Computing Center, a DOE Office of Science User Facility under the same contract. Additional support for DESI was provided by the U.S. National Science Foundation (NSF), Division of Astronomical Sciences under Contract No. AST-0950945 to the NSF’s National Optical-Infrared Astronomy Research Laboratory; the Science and Technology Facilities Council of the United Kingdom; the Gordon and Betty Moore Foundation; the Heising-Simons Foundation; the French Alternative Energies and Atomic Energy Commission (CEA); the National Council of Humanities, Science and Technology of Mexico (CONAHCYT); the Ministry of Science and Innovation of Spain (MICINN), and by the DESI Member Institutions: www.desi.lbl.gov/collaborating-institutions. The DESI collaboration is honored to be permitted to conduct scientific research on I’oligam Du’ag (Kitt Peak), a mountain with particular significance to the Tohono O’odham Nation. Any opinions, findings, and conclusions or recommendations expressed in this material are those of the author(s) and do not necessarily reflect the views of the U.S. National Science Foundation, the U.S. Department of Energy, or any of the listed funding agencies.

The Hyper Suprime-Cam (HSC) collaboration includes the astronomical communities of Japan and Taiwan, and Princeton University. The HSC instrumentation and software were developed by the National Astronomical Observatory of Japan (NAOJ), the Kavli Institute for the Physics and Mathematics of the Universe (Kavli IPMU), the University of Tokyo, the High Energy Accelerator Research Organization (KEK), the Academia Sinica Institute for Astronomy and Astrophysics in Taiwan (ASIAA), and Princeton University. Funding was contributed by the FIRST program from the Japanese Cabinet Office, the Ministry of Education, Culture, Sports, Science and Technology (MEXT), the Japan Society for the Promotion of Science (JSPS), Japan Science and Technology Agency (JST), the Toray Science Foundation, NAOJ, Kavli IPMU, KEK, ASIAA, and Princeton University. This paper makes use of software developed for the Large Synoptic Survey Telescope. We thank the LSST Project for making their code available as free software at  http://dm.lsst.org. This paper is based [in part] on data collected at the Subaru Telescope and retrieved from the HSC data archive system, which is operated by the Subaru Telescope and Astronomy Data Center (ADC) at National Astronomical Observatory of Japan. Data analysis was in part carried out with the cooperation of Center for Computational Astrophysics (CfCA), National Astronomical Observatory of Japan. The Subaru Telescope is honored and grateful for the opportunity of observing the Universe from Maunakea, which has the cultural, historical and natural significance in Hawaii. 
\end{ack}

\bibliographystyle{abbrvnat}
\bibliography{references2}







\end{document}